\documentclass[11pt]{article}
\usepackage{graphicx}
\usepackage{cs12,html,hyperref}

\begin{document}
 
\title{X-ray Spectra and Light Curves of AR Lac: Temperature Structure,
  Abundances, and Variability.
  }
 
\author{David Huenemoerder\altaffilmark{1}, Claude Canizares\altaffilmark{1}, 
Kevin Tibbetts\altaffilmark{1}}
\altaffiltext{1}{Massachusetts Institute of Technology}

\index{*AR Lac}
\index{Chandra}
\index{spectra, X-ray}

\begin{abstract}

We observed AR Lac, an eclipsing RS CVn binary star, with the Chandra
High Energy Grating Spectrometer for a total of 100 ks divided into
six intervals covering both quadratures and eclipses. We repeated
observations at each phase.  At least two flares were seen, in which
the flux increased by factors of two and four.  The flares occurred
near eclipse phases and compromised detection of eclipse modulation.
The quadrature fluxes were the most stable, but they also show
non-repeating trends.  Quadrature line profiles are broadened relative
to eclipse profiles, presumably due to orbital velocity Doppler
shifted emission from each binary component.  The spectrum appears to
be iron poor and neon rich, similar to HR 1099, but not as extreme as
II Pegasi.  We will examine line strengths, widths, and positions
vs. phase, and present preliminary differential emission measure
models and abundance determinations using the APED emissivity
database.

This work is supported by NASA contract NAS8-38249 (HETG) and SAO
SV1-61010 (CXC) to MIT.

\end{abstract}

\section{Introduction}
RS CVn binaries were first noted for their sinusoidal optical light
curve variability, interpreted as large scale analogs of sunspots
(Hall, 1976).  Subsequent studies showed them to have optical
(hydrogen and calcium; Huenemoerder \& Ramsey 1984) and ultraviolet
(carbon, silicon, oxygen, nitrogen; Simon and Linsky 1980) emission
lines also analogous to the Solar chromosphere and transition region.
They were later found to be luminous in X-rays (Walter \& Bowyer,
1981), extending the Solar analogy to the corona, but with 1000 times
the activity level.  Eclipsing systems like AR Lac permit us to study
coronal phenomena for stars of known mass, size, and rotational
velocity.  Emission line spectroscopy provides temperature and density
diagnostics, which together with geometrical constraints, can provide
insights regarding sizes of magnetic loops.  Eclipse and rotational
modulation provide further information on the size and distribution of
the emitting structures.  Together in the context provided by other
stars and long term trends, we hope to learn about the fundamental
physical processes which heat stellar coronae.

We observed AR Lac with the Chandra High Energy Grating Spectrometer
(HETGS) for 97 ks, divided into 6 separate observations which sampled
eclipse and quadrature phases.  The HETGS has a maximum spectral
resolution ($\lambda/FWHM$) of about 1500, covering the range of 1.8
to 30 \AA\ with two complementary gratings, the Medium, and the High
Energy Gratings (MEG and HEG, respectively).  Approximately 20 years
prior to these Chandra observations, Walter, Gibson, and Basri (1983)
mapped X-ray emission with the {\em Einstein} observatory, and found
both extended and compact emission regions with an asymmetric
distribution.  Neff et al. (1989) mapped the chromosphere using
ultraviolet lines.  We now have high resolution spectra which resolve
the emission into a multitude of emission lines from highly ionized
elements.

\section{Light Curve}

\vspace{-0.5cm}
\begin{figure}[htb]
  \centerline{\includegraphics*[width=15cm]{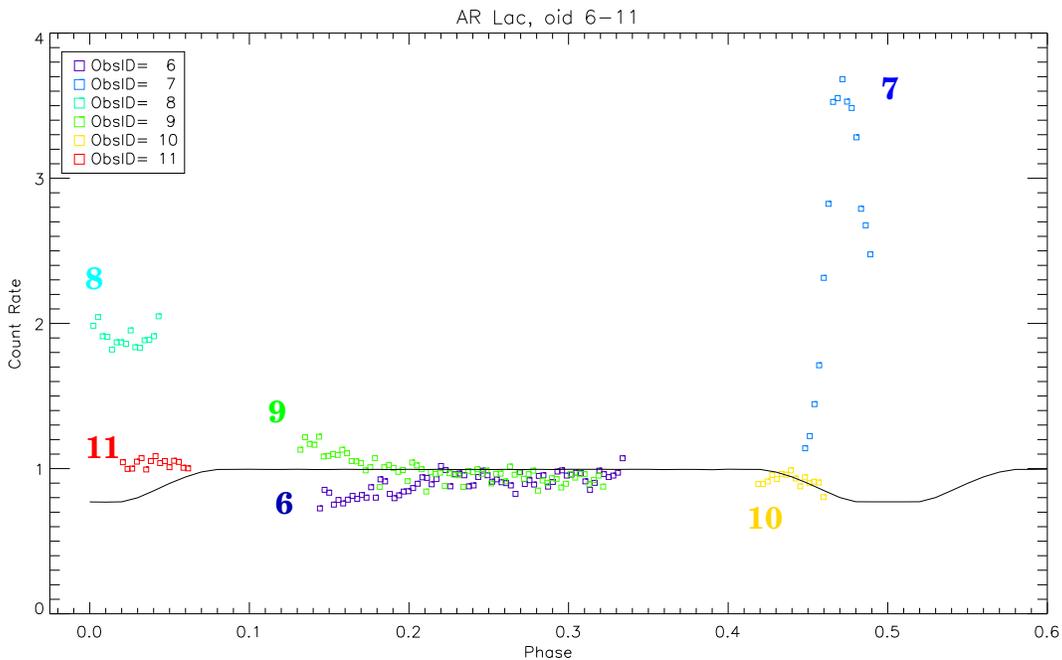}}
  \caption{\sf
    The count-rate is shown against orbital phase.  Each observation
    ID is color coded and labeled. The time-order of ObsIDs is 6, 8,
    7, 9, 11, 10. One-sigma uncertainties are slightly larger than the
    symbol size.  The solid line is a simple light curve model
    assuming uniformly emitting disks of equal intensity.  It was
    arbitrarily placed at a level which suggests that ObsID 10 may show
    eclipse modulation as the larger K0 IV occults the G2
    IV.}
  \label{fg:lcurve}
\end{figure}
A light curve was made by binning diffracted photons in orders -3 to
+3 (excluding 0) in both HEG and MEG arms.  Figure~\ref{fg:lcurve}
shows the count-rate against orbital phase, which is also rotational
phase, since the system is tidally synchronized (the period of AR Lac
is 1.98 days).  ObsID 7 shows a strong flare, caught during the rise
and part of the decay. ObsID 8 has a high rate, but is not obviously
rising or falling.  The light curve shows no primary eclipse,
indicating that the G2 IV star (smaller component) is X-ray dark, at
least on one hemisphere.

\section{Doppler Shifts}

\begin{figure}[htb]
\centerline{\includegraphics*[width=11cm]{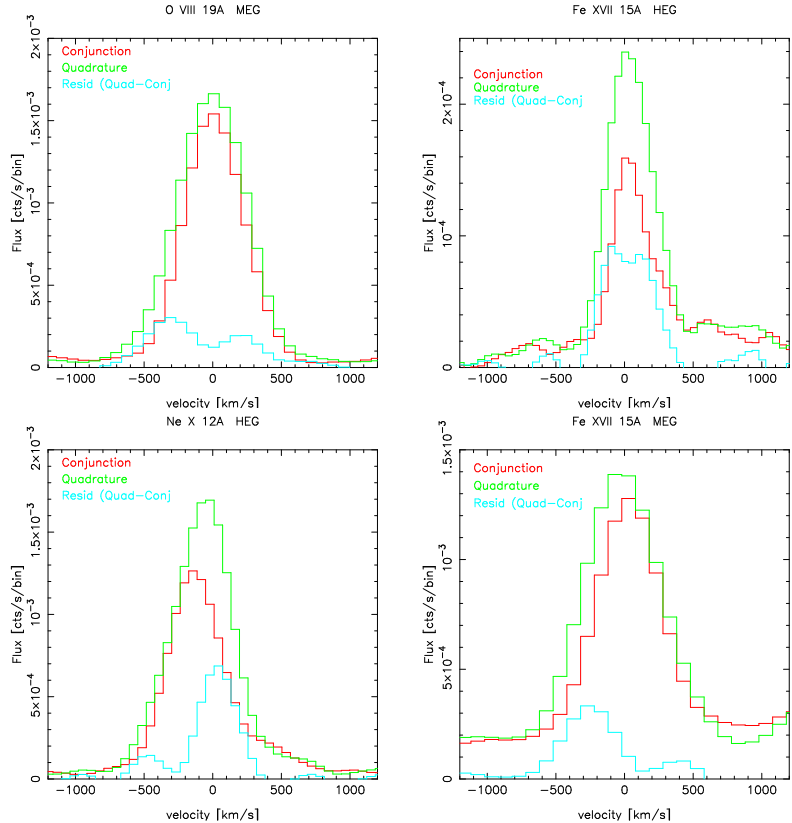}}
\caption{\sf
  A comparison of line profiles between quadrature (green) and 
  conjunction (red) for wavelengths having the highest spectral 
  resolution. Residuals are shown in light blue.  } 
\label{fg:doppler}
\end{figure}
The orbital radial velocity amplitude of AR Lac is 230 km/s.  Solar
flares have been seen to have velocity fields of up to 400 km/s.  In
Figure~\ref{fg:doppler}, we compare line profiles for the quadrature
times (ObsIDs 9 and 6) to profiles during times of conjunction, which
also coincidentally have flares (ObsIDs 7,8,10, \& 11). We have looked
at lines near the highest instrumental resolution: O~{\sc viii} 19\AA\ 
MEG, Ne~{\sc x} 12 \AA\ HEG (blended w/ Fe~{\sc xvii}), and Fe~{\sc
  xvii} 15\AA\ (HEG and MEG).  It
is clear that lines during quadrature are broader than during
conjunction, which is expected if both components are X-ray bright.
Future work will examine line centroids versus time versus order to
better quantify the origin of the broadening and shifts.

\section{Density Diagnostics}

\begin{figure}[htb]
\centerline{\includegraphics*[width=10cm]{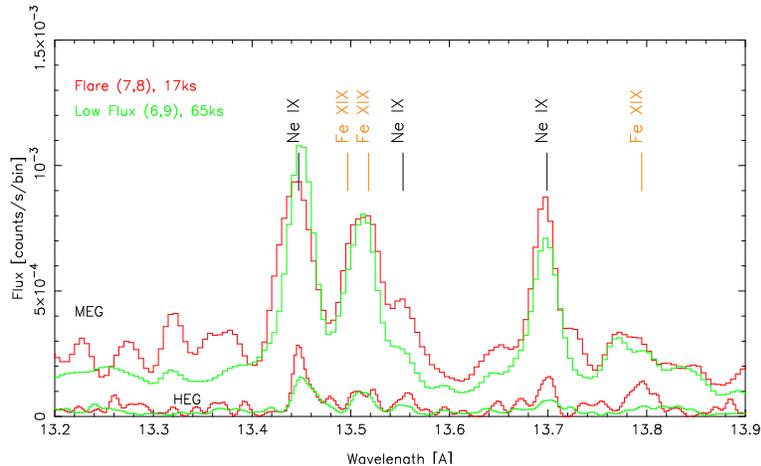}}
\caption{\sf
  The Ne~{\sc ix} He-like triplet region for both MEG (upper
  traces), and HEG (lower traces). HEG has higher resolution, but less
  effective area.  The spectra from the flare states are shown in red,
  and the low-flux states are in green.  The forbidden line (13.7 \AA)
  to intercombination line (13.55 \AA) ratio is density
  sensitive.}\label{fg:density}
\end{figure}

The helium-like triplet lines provide density diagnostics which are
independent of relative abundance ratios and which are relatively
independent of temperature.  The Ne~{\sc ix} region is shown in
Figure~\ref{fg:density} for flare and low-flux intervals. The
forbidden line (13.7 \AA) to intercombination line (13.55 \AA) ratio
is density sensitive and decreases for densities above about $\log
N_e=12$ (cgs).  It appears likely that the ratio decreases during
flares.

\section{The Full Spectrum}

\begin{figure}[htb]
  \centerline{\includegraphics*[width=8cm]{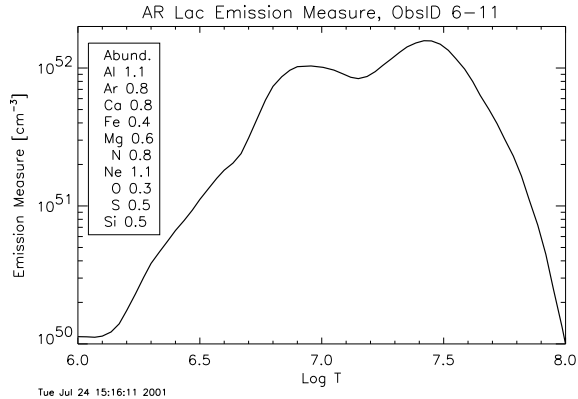}}
  \caption{\sf
    A provisional differential emission measure derived from
    line fluxes integrated over all observations. This DEM was used to
    generate the model spectrum shown in Figure~\ref{fg:spec}.}\label{fg:DEM}
\end{figure}

We have derived a provisional differential differential emission
measure model by measuring line fluxes in the total integrated
spectrum, and then fitting to them a differential emission measure
(DEM) with variable abundances.  The model is considered as
provisional because of the great variability of the light curve and
because the system is composite.  After an initial temperature
distribution was derived, the DEM and abundances were scaled to give
an approximate match to line-to-continuum ratios.  The fit is not
unique, and we will be fitting shorter, more stable intervals of the
spectrum.  At low temperatures, the fit is unconstrained because there
are no lines from plasma cooler than about $\log T = 6.3$.  The hot
end is constrained by emission from high ionization stages of Fe and
by the continuum.  The provisional DEM is shown in
Figure~\ref{fg:DEM}.

\begin{figure}[htb]
  \centerline{\includegraphics*[width=12cm]{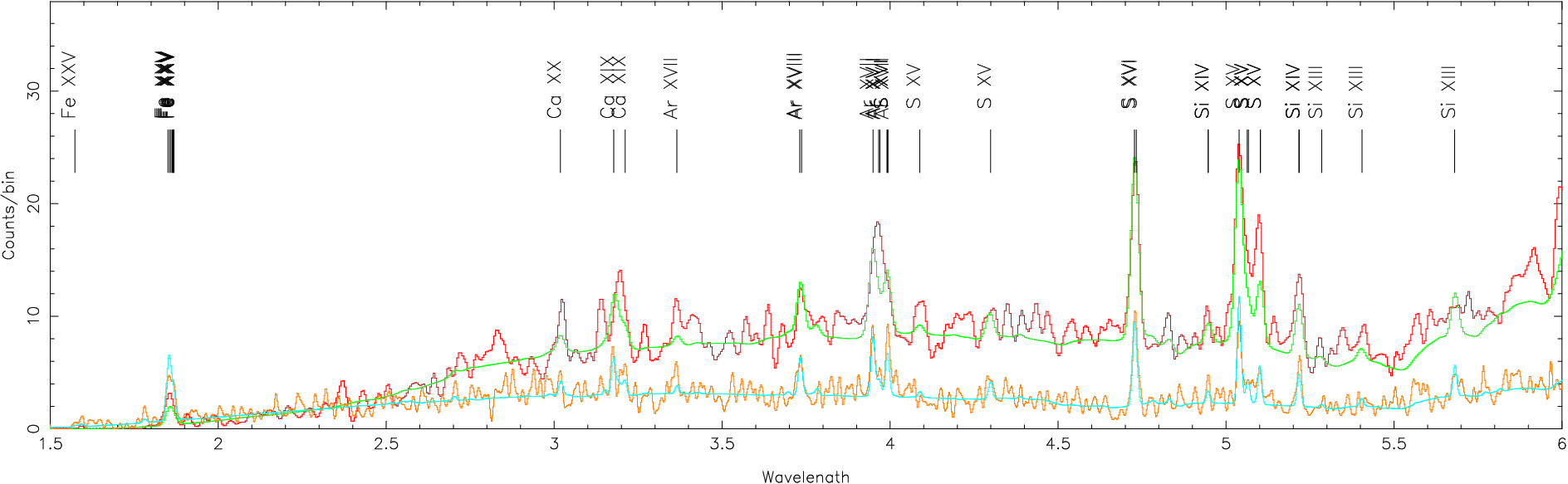}}
  \centerline{\includegraphics*[width=12cm]{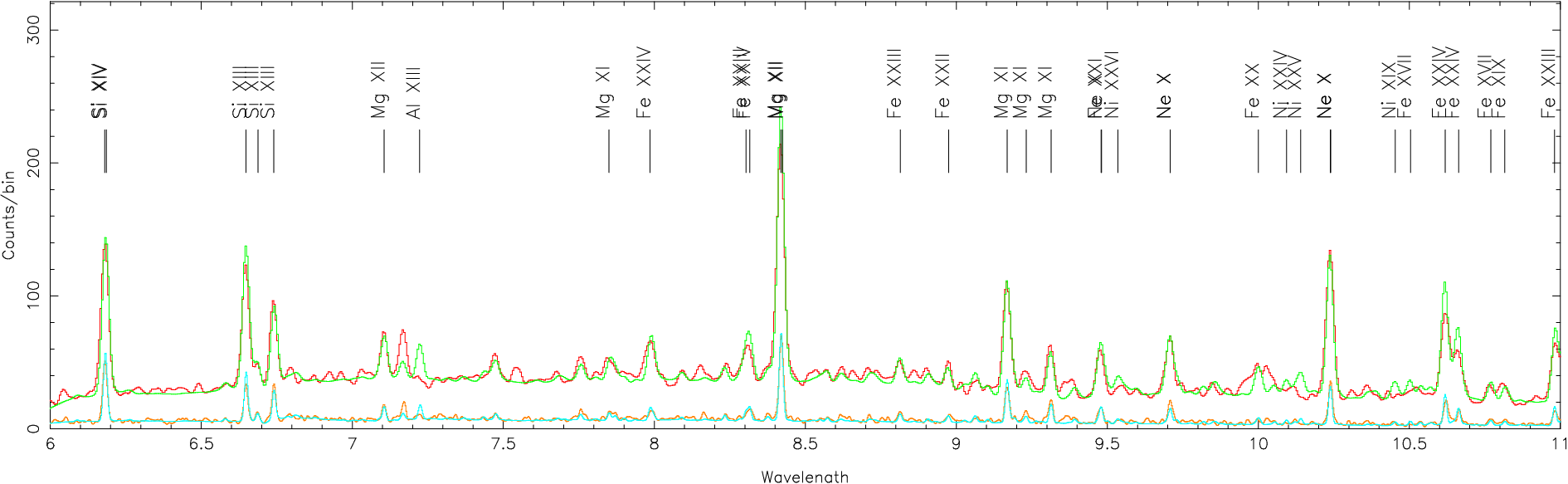}}
  \centerline{\includegraphics*[width=12cm]{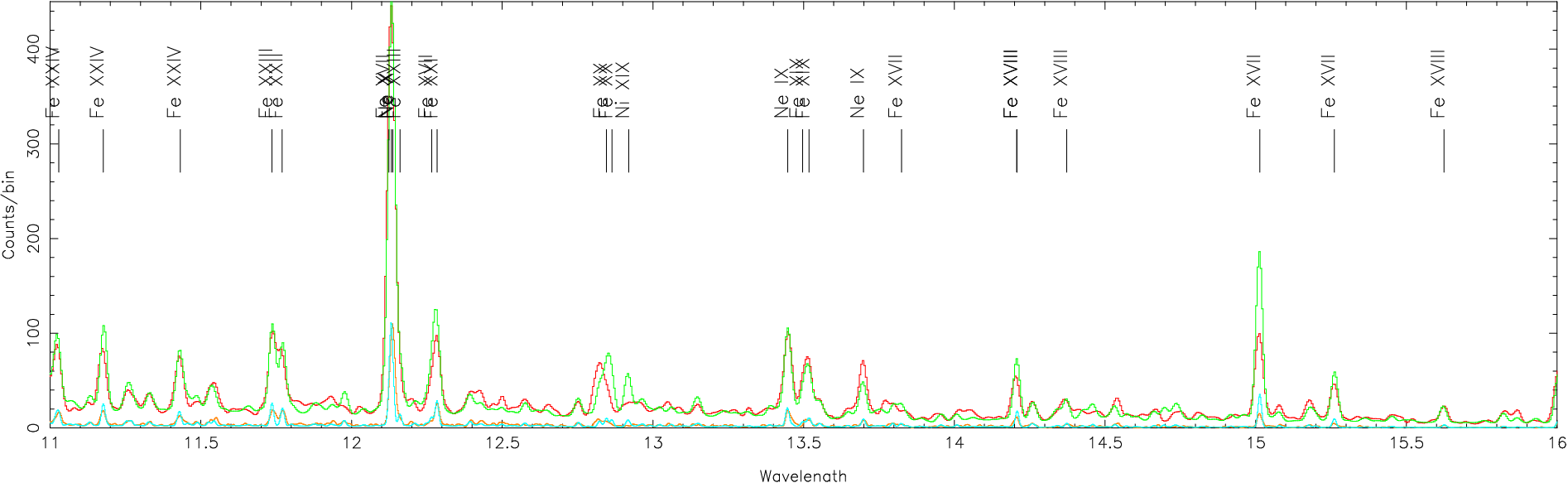}}
  \centerline{\includegraphics*[width=12cm]{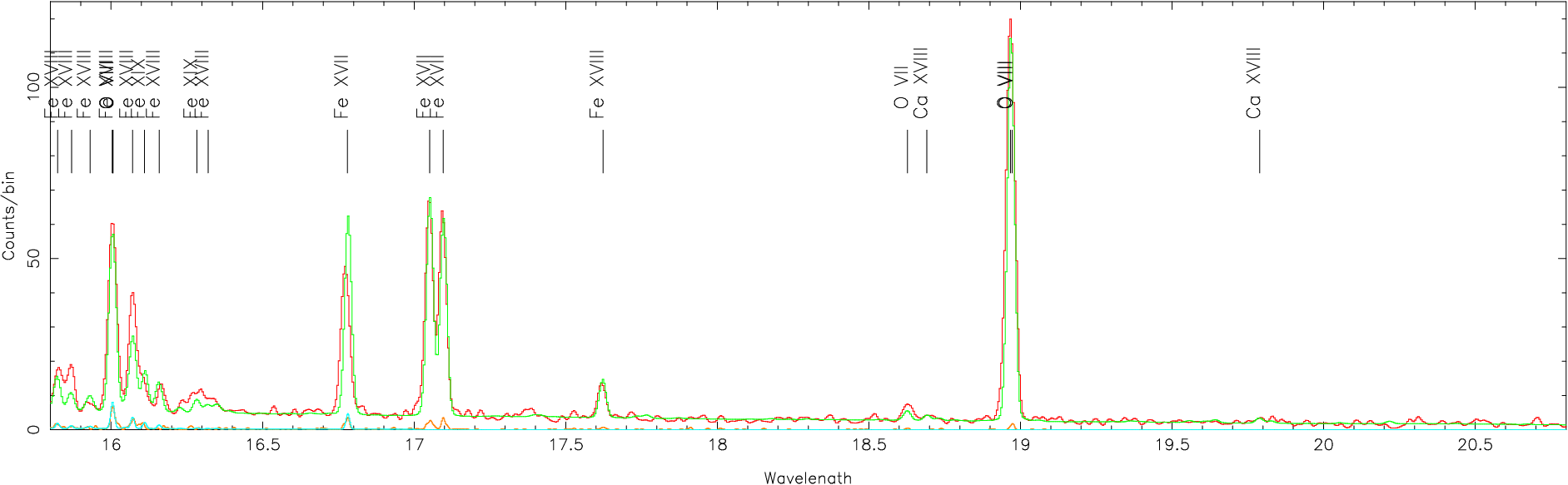}}
  \centerline{\includegraphics*[width=12cm]{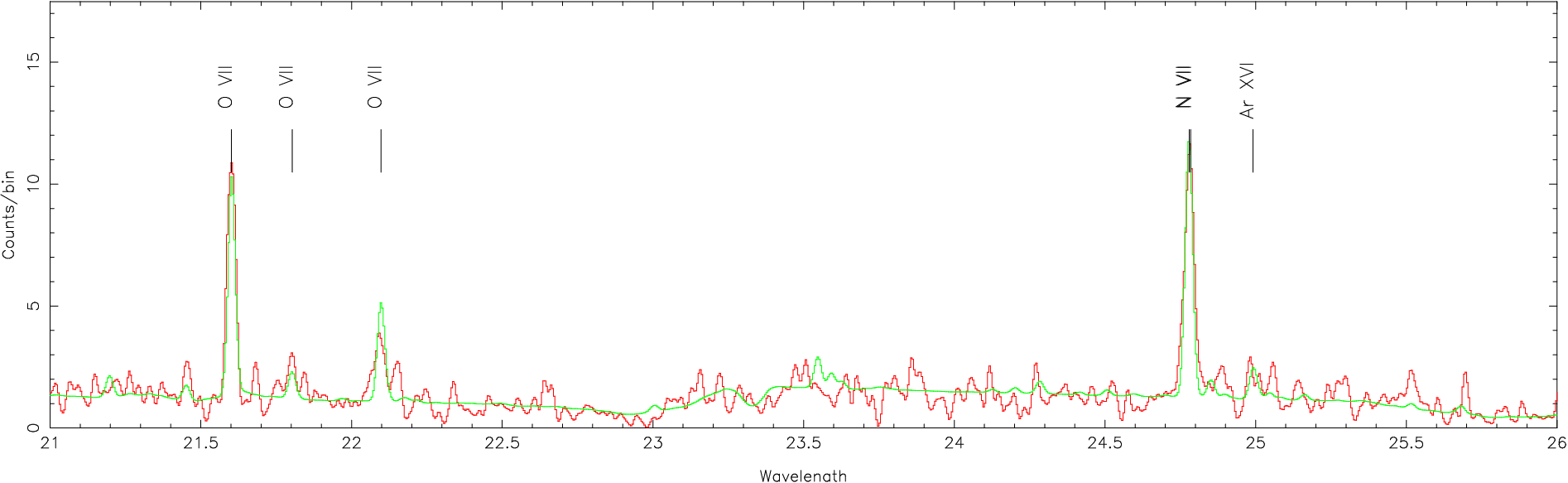}}
  \caption{\sf
    The Chandra HETGS spectrum of AR Lac. The MEG spectrum is red, and
    the HEG orange.  The MEG model is green, and HEG model blue.
    Since the MEG has more effective area, it is the upper curve.  The
    HEG has higher resolution. Each panel has its own vertical
    scale.}\label{fg:spec}
\end{figure}

Counts spectra are extracted for the entire 97 ks observation (ObsIDs
6-11), which is shown in Figure~\ref{fg:spec}.  Plus and minus 1st
orders were summed and smoothed by 0.008 \AA\ (Medium Energy Grating,
MEG) or 0.004 \AA\ (High Energy Grating, HEG) for display.  The marked
lines are the brightest features for the adopted model.  The model was
binned from the APED (Astrophysical Plasma Emission Database; Smith et
al 2001) and folded through the instrumental response.

\section{Conclusions}

The HETGS spectra show that the coronae of AR Lac are highly variable,
present on both stellar components, and asymmetrically distributed.
Ionic species detected indicate a broad range in plasma temperatures.
Relative abundances are subject to further more detailed analysis, but
preliminary results fairly reliably indicate that iron is depleted by
about a factor of three relative to neon, based upon Ne~{\sc ix} and
Fe~{\sc xvii} ratios, since these ions have similar emissivity
functions with temperature.  We also detected a possible density
enhancement during the flare state.

\acknowledgments

This work is supported by NASA contract NAS8-38249 (HETG) and SAO
SV1-61010 (CXC) to MIT.

\index{*AR Lac}

\end{document}